\documentclass[superscriptaddress]{revtex4}

\usepackage{graphicx}

\begin{document}

\title{Long Brief Pulse Method for Pulse-wave \\ modified Electroconvulsive Therapy}
\author{Hiroaki Inomata}
\email{inomatah0612@gmail.com}
\affiliation{Tokyo Metropolitan Matsuzawa Hospital}
\affiliation{Yokohama City University}
\affiliation{Tokyo Metropolitan Institue of Medical Science}
\author{Hirohiko Harima}
\affiliation{Tokyo Metropolitan Matsuzawa Hospital}
\author{Masanari Itokawa}
\affiliation{Tokyo Metropolitan Matsuzawa Hospital}
\affiliation{Tokyo Metropolitan Institue of Medical Science}
%\date{\today}

\begin{abstract}
Modified-Electroconvulsive Therapy (m-ECT) is administered for the treatment of various psychiatric disorders. The Seizure Generalization Hypothesis holds that propagation of the induced seizure throughout the whole brain is essential for the effective ECT intervention. However, we encounter many clinical cases where, due to high thresholds, seizure is not induced by the maximum dose of electrical charge.  Some studies have indicated that the ultrabrief pulse method, in which pulse width is less than 0.5millisecond (ms), is more effective at inducing seizure than conventional brief pulse (0.5ms-2.0ms). Contrary to the studies, we experienced a case of schizophrenia in which m-ECT with 1.0 and 1.5 ms width pulse (referred to as ‘long’ brief pulse as 0.5ms width pulse is the default in Japan) succeeded in inducing seizure, whereas ultrabrief pulse failed to induce seizure. This case is described in detail. Moreover, we discuss the underlying mechanism of this phenomenon.
\end{abstract}
\maketitle

\section*{Introduction}
Modified-Electroconvulsive Therapy (m-ECT) is administered for the treatment of various psychiatric disorders. The Seizure Generalization Hypothesis, which underlies the mechanism of ECT, holds that propagation of induced seizure throughout the entire brain is essential for effective ECT intervention\cite{ECTbook} . However, there are many clinical cases where, due to high thresholds, seizure is not induced by the maximum dose of electrical charge.  In these cases, the following procedures are considered options for inducing seizure; (i) using the older method of sine-wave ECT (ii) promoting hyperventilation in patients \cite{LooAug} (iii) using anesthetic agents such as ketamine with ECT \cite{LooAug}.  However, these are not standard methods as sine-wave ECT induces more severe side effects than pulse-wave ECT, and not all anesthesiologists are fully trained in the latter two procedures.

Recently randomized control trials focusing on pulse width have been conducted\cite{Sackeim,Loo}.  Sackeim et al \cite{Sackeim} reported that the ultrabrief pulse method, in which pulse width is less than 0.3millisecond (ms), induces more therapeutic effects and fewer side effects and requires less electrical charge to induce seizure compared to conventional brief pulse (1.5ms).

It could be predicted then that the ultrabrief pulse would be more effective in inducing seizure in patients with high thresholds.  Contrary to this, we experienced a case of schizophrenia in which m-ECT with 1.0 and 1.5ms width pulse (referred to as ‘long’ brief pulses as 0.5ms width pulse is the default in Japan) succeeded in inducing seizure, whereas ultrabrief pulse failed. We present this case in detail and discuss the possible underlying mechanisms.

Written informed consent was obtained from the patient and his wife (legal guardian). All personal information has been anonymized. 

\section*{Case presentation}
The patient is a 35-year-old schizophrenic Japanese male. His history of illness started at age 23 with symptoms of auditory hallucinations, persecutory delusions and psychomotor excitement. Subsequently these symptoms relapsed every one to three years.  He was discharged from hospital three years ago and was being followed as an outpatient. One month ago, he complained to his doctor that he was ‘being watched by strangers’.  The atypical antipsychotic blonanserine and the mood stabilizer valproate were added to his medication but these produced the persecutory delusion that the new therapy was part of an experiment instigated by his doctor. He became aggressive with verbal threats and was admitted to our hospital due to a lack of available beds in his regular treating hospital. 

On admission, he was psychotic and extremely agitated, and needed to be restrained. He initially refused treatment but eventually agreed to his old treatment regime excluding blonanserine and valproate. His medical history revealed that medication was of little effect for his relapsed symptoms while m-ECT was effective, although sine-wave ECT was necessary due to his high threshold. 

The presence of brain disease such as tumours was excluded after review of Computed-Tomography images. ECT treatment was instigated three days post admission.  For the ECT interventions, we used a Somatic Thymatron ECT device, placing electrodes in a bitemporal configuration and administering propofol at 1.0mg/kg as an anesthetic induction and succinylcholin at 1.0 mg/kg as a muscle relaxant. 

Figure 1 shows the clinical course of this patient.  In trials 1 to 3, we used a ‘LOW 0.5’ setting, in which the pulse width is fixed at 0.5ms, but this failed to induce seizure at the maximum dose of electrical charge (504 milicoulomb).  We then changed the Thymatron setting from ‘LOW 0.5’ to ‘LOW 0.25’, with pulse width fixed at 0.25 ms, for trial 4, performed immediately after trial 3.  Contrary to expectations, this did not induce seizure. In trial 5, a setting of ‘LOW 0.5’ succeeded in inducing 9-second seizure, recognized by EEG and EMG charts. This seizure, however, was deemed insufficient due to its short duration time.  We then changed pulse width settings to 1.0ms and administered trial 6.  This setting successfully induced seizure with the desirable waveform and of sufficient duration (Fig 2).

For the remainder of the treatment we administered ECT, decreasing the electrical charge to avoid side effects. At the 1.0ms pulse width, we achieved desirable seizures with 60 percent of the maximal charge.  At the 1.5ms pulse width, we succeeded in inducing seizure at 40 percent of the maximal charge.  We stopped ECT treatment after 11 trials as his psychiatric condition had improved considerably from the day of admission.  He was then transferred to a chronic ward and prepared for discharge.

\section*{Discussion}
For this patient, all trials at 0.25ms width pulse failed to induce seizures while 0.5ms width pulse was successful only once (in trial 5).  It is likely that falling serum valproate concentrations enabled seizure induction in trial 5 whereas the same pulse width failed to induce convulsions in the earlier trials (1 through 3).  The 1.0ms width pulse succeeded in inducing therapeutic seizure with desirable waveform, while the 1.5ms width pulse trials induced seizure of unknown clinical effect.

In summary, long brief pulse was more effective for inducing seizure than ultrabrief pulse for this patient.  Taken together with recent RCT studies, this case suggests that seizure threshold depends on pulse width. However, results are contradictory. In the RCT studies, ultrabrief pulse was more effective than long brief pulse for inducing seizures but, in this case, the reverse was true. 

Recent RCT studies of ultrabrief pulses are based on the electrophysiological fact that the chronaxie (the most effective pulse width in firing neuron) of neurons in the mammalian central nervous system lies within 0.1-0.3ms \cite{Peterchev}.   However, West et al reported that the strength-duration curve of one-thirds of neurons is right-shifted, even in normal subjects, so that their chronaxie is prolonged \cite{West}.  We speculate that in our patient the strength-duration curve involved in ECT-induced seizures might be right-shifted resulting in prolonged chronaxie through to about 1.0ms. Figure 3 shows a possible mechanistic explanation for this apparent contradiction. Moreover, this view accords well with the fact that there is little clinical difference between a 0.25ms and 0.5ms width pulse \cite{Niemantsverdriet}. If the strength-duration curve is not right-shifted (in an individual, or where sample size is such that this can be assumed), then the difference in electrophysiological response produced by a pulse width of a 0.25ms compared to 0.50ms is negligible relative to the difference in response between 0.30ms compared to 1.5ms.

Although our hypothesis is likely to bridge clinical usage of ECT and relevant fundamental research, it is first necessary to confirm whether our observations are limited to this case or are applicable to a wider group of patients with careful ECT treatment.

\section*{Acknowledgments}
We thank the medical staff at Matsuzawa Hospital especially Dr Okazaki.

\clearpage{}
\section*{Figure}

\begin{figure}[h]
\includegraphics[width=16cm]{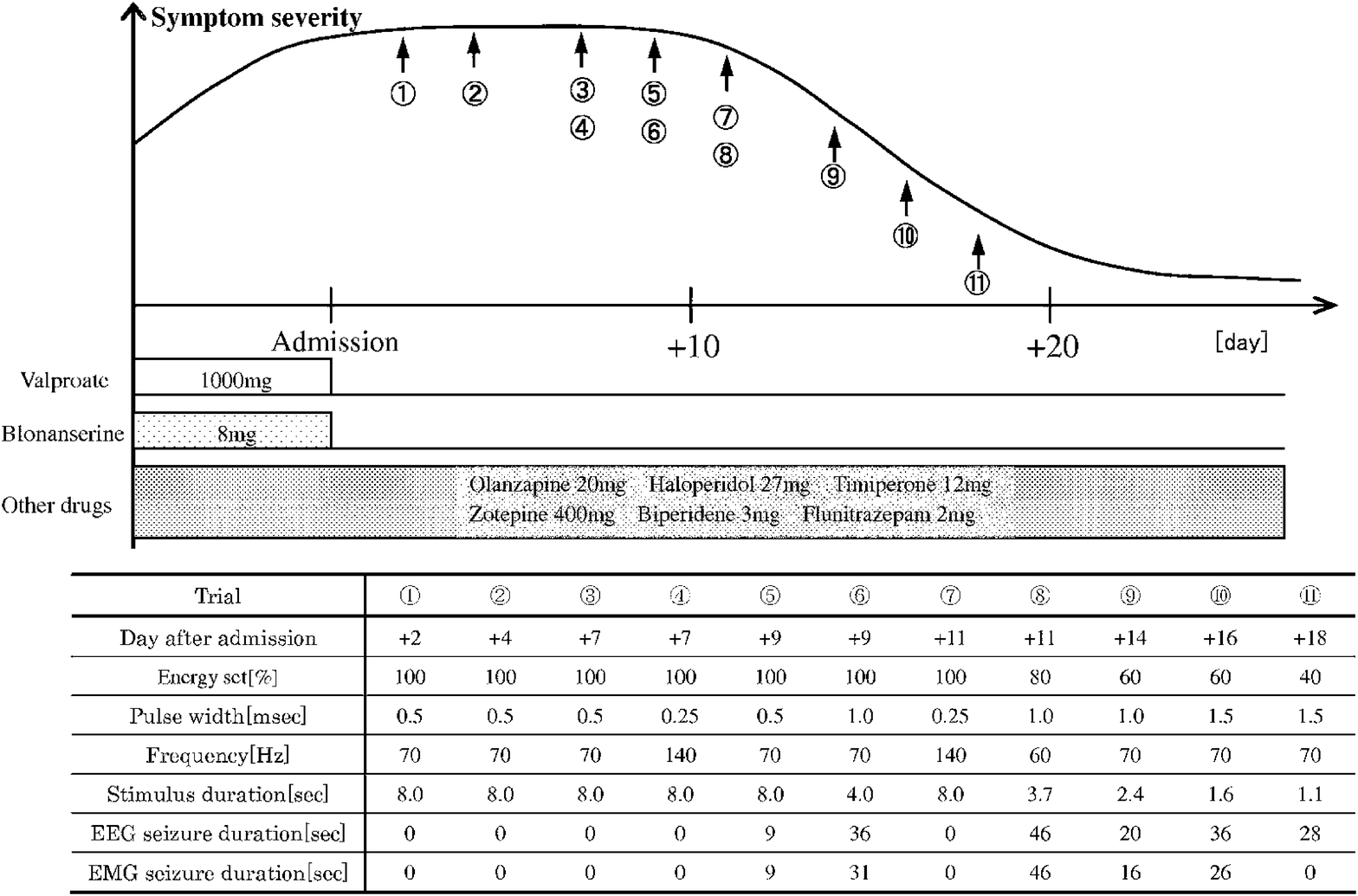}
\caption{{\bf Clinical course}}
\label{figClinicalCourse}
\end{figure}

\begin{figure}[h]
\includegraphics[width=17cm]{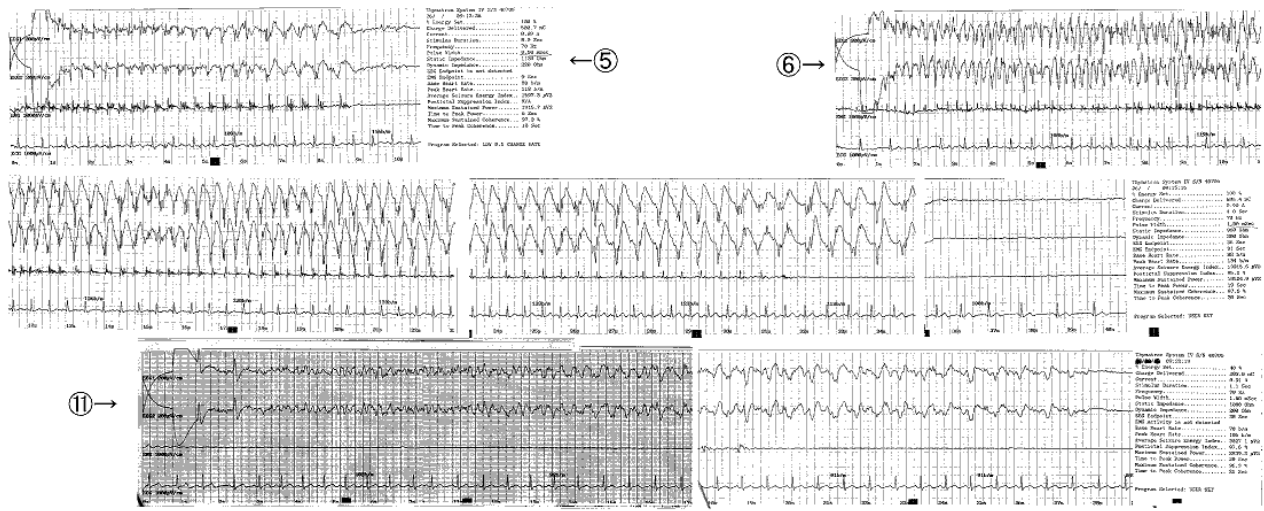}
\caption{{\bf EEG and EMG charts of the 5th, 6th, and 11th trials}}
\label{figEEGandEMG}
\end{figure}

\begin{figure}[h]
\includegraphics[width=18cm]{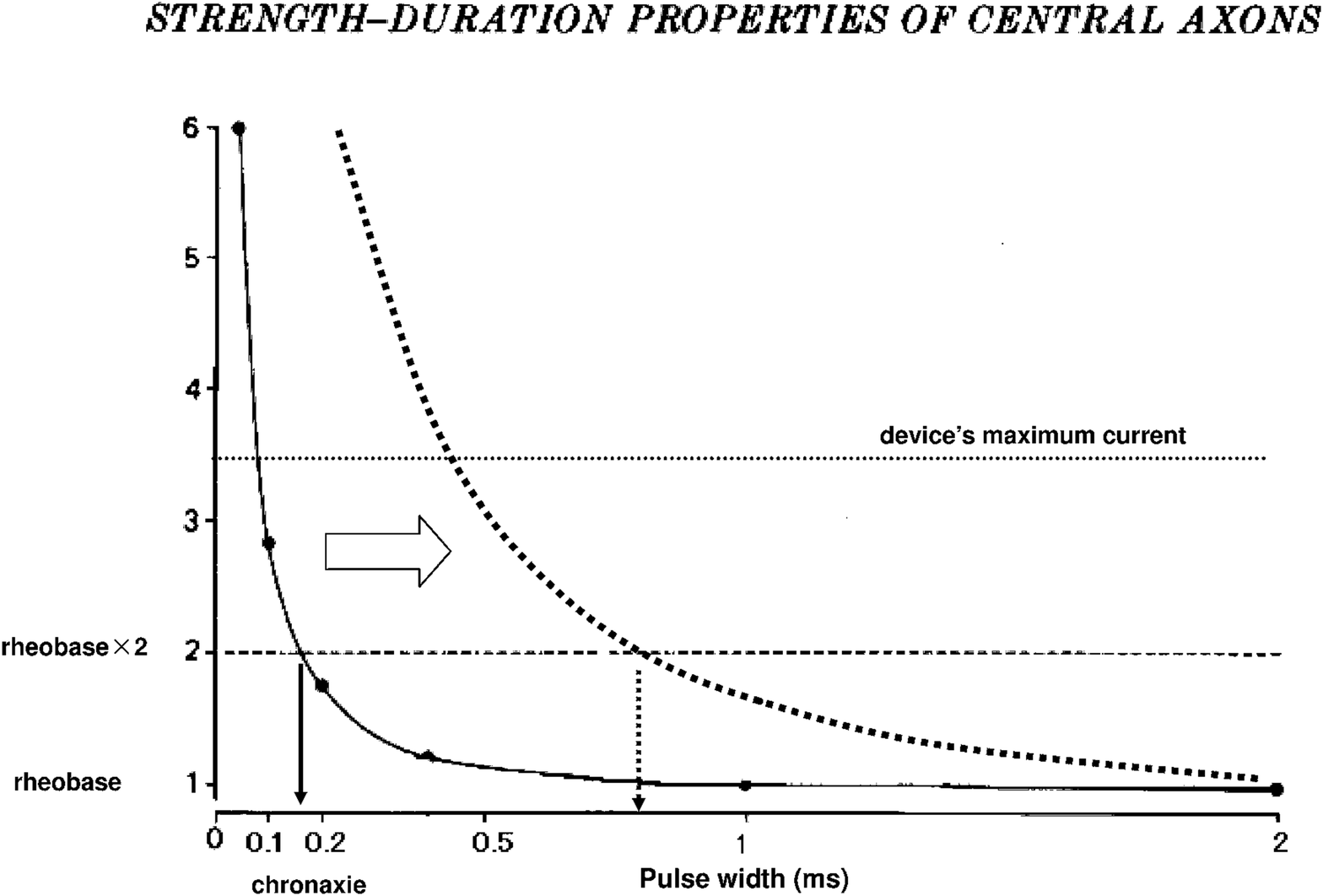}
\caption{{\bf Strength-duration curve} This graph is modified from West et al\cite{West}}
\label{figRight-shifted}
\end{figure}

\end{document}